\documentclass[sigconf]{acmart}

\AtBeginDocument{%
  }

\usepackage{algorithm}
\usepackage{algorithmic}

\DeclareRobustCommand{\cal}[1]{\ensuremath{\mathcal{#1}}}

\DeclareRobustCommand{\E}[0]{\mathbb{E}}

\DeclareRobustCommand{\R}[0]{\mathbb{R}}

\DeclareRobustCommand{\eqlabel}[1]{\label{eq:#1}}                                     	    
\DeclareRobustCommand{\eq}[1]{\begin{equation}\eqlabel{#1}}                           

\DeclareRobustCommand{\eqend}{\end{equation}}                                         	    
\DeclareRobustCommand{\eqref}[1]{\mbox{(\ref{eq:#1})}}                                   

\DeclareRobustCommand{\eqary}{\begin{eqnarray*}}
\DeclareRobustCommand{\eqaryend}{\end{eqnarray*}}

\usepackage{caption} 
\usepackage{array,ragged2e}
\newcolumntype{P}[1]{>{\RaggedRight\arraybackslash}p{#1}}

\newcommand{\minus}{\scalebox{0.75}[1.0]{$-$}}

\setcopyright{acmcopyright}
\copyrightyear{2022}
\acmYear{2022}

\acmConference[ICAIF '22]{3rd ACM International Conference on AI in Finance}{November 2--4, 2022}{NY}

\begin{document}

\title{Deep Hedging: Continuous Reinforcement Learning for Hedging of General Portfolios across Multiple Risk Aversions}

\author{Phillip Murray}
\affiliation{%
  \institution{J.P. Morgan}
  \streetaddress{}
  \city{London}
  \country{UK}
}

\author{Ben Wood}
\affiliation{%
  \institution{J.P. Morgan}
  \streetaddress{}
  \city{London}
  \country{UK}
}

\author{Hans Buehler}
\affiliation{%
  \institution{TU Munich}
  \streetaddress{}
  \city{Munich}
  \country{Germany}
}
\email{}

\author{Magnus Wiese}
\affiliation{%
  \institution{TU Kaiserslautern}
  \streetaddress{}
  \city{Kaiserslautern}
  \country{Germany}
}
\email{}

\author{Mikko S. Pakkanen}
\affiliation{%
  \institution{Imperial College London}
  \streetaddress{}
  \city{London}
  \country{UK}
}

\renewcommand{\shortauthors}{Murray et al.}

\begin{abstract}
We present a method for finding optimal hedging policies for arbitrary initial portfolios and market states.
We develop a novel actor-critic algorithm for solving general risk-averse stochastic control problems and use it to learn 
hedging strategies across multiple risk aversion levels simultaneously. We demonstrate the effectiveness of the approach 
with a numerical example in a stochastic volatility environment.

\end{abstract}

\ccsdesc[500]{Theory of computation~Sequential decision making}
\ccsdesc[500]{Theory of computation~Reinforcement learning}

\keywords{Deep Hedging, Reinforcement Learning, Risk Averse, Transaction Costs
}


\maketitle

\section{Introduction}

Traditionally, financial inventory management of securities and over-the-counter (OTC) derivatives
has relied on classic \emph{complete market} models of quantitative finance, such as the seminal work by Black, Scholes and Merton.
There, we assume that all risks can be eliminated by cost-free continuous-time trading in markets
with infinite depth, whose price processes evolve undisturbed by our trading activity.
The main characteristic of this approach is that we use 
the sensitivities of the complete-market value of our inventory with respect to changes
in market parameters --
\emph{the Greeks} -- as risk management signals.

However, since the idealized assumptions of a complete market do not
apply in real markets, it is not surprising that complete market models require constant manual
adjustments and oversight, for example adjusting delta by ``skew delta", smoothing barriers
priced with local volatility, and taking into account market impact when trading vega to hedge auto-callable
products.
Financial inventory management is therefore
to this day heavily dependent on human decision making and heuristic adjustments.

This approach made sense when availability of data and computation power was scarce. However, in recent years
there has been a growing interest in the application of modern machine learning methods to the pricing and hedging of
portfolios of OTC derivatives. In particular, the approach known as \emph{deep hedging} \cite{DH} has paved the way
to a model-agnostic, data-driven framework for scalable decision making in financial inventory management. 

In the deep hedging framework, it is possible to construct a dynamic risk-adjusted hedging strategy in
complex hedging instruments under market frictions for a fixed portfolio with a fixed time horizon~$T$ using
episodic policy search, using market simulation until the maturity of the portfolio. Further iterations of the deep hedging algorithm have been explored, such as applications to rough volatility \cite{horvath2021deep}, 
and American-style options \cite{becker2020pricing}.

However, to date, all approaches face the significant limitation that the model must be trained using a fixed portfolio, 
over a fixed time horizon. Thus, we need to train a new model every time we do a new trade. Furthermore, the model is trained against some risk-averse objective, often with 
a somewhat arbitrarily chosen risk preference. Hence, in market states where we might want to change our tolerance to risk, such 
as times of high volatility, we again need to train a new model with this new risk aversion level.

In this article, we address all of the above challenges. Firstly, we learn general policies for arbitrary portfolios by
employing a common feature representation of our portfolio instruments. Specifically, we mimic exactly what a human trader 
does - we compute the ``book value", risk metrics, and \emph{Greeks} of our portfolio at each market state. This therefore gives us a 
feature space which is shared across different payoffs, and is also a \emph{Linear Markov Representation (LMR)} in the sense that we
can represent linear combinations of portfolio instruments by the corresponding linear combination of their feature vectors. By doing so, we 
learn a policy which is a mapping from this feature space to actions, which learns the correct state-dependent adjustment to model sensitivities.   

Secondly, we formulate deep hedging in a more
classical reinforcement learning framework, and derive a \emph{risk-averse} Bellman representation of the deep 
hedging problem. We then develop a novel actor-critic algorithm for solving this risk-averse problem. Our model-free, on-policy 
algorithm uses deep function approximators to learn both policies and value functions for general portfolios of instruments. By 
using monetary utility functions, our value function also represents an indifference price for the portfolio. Hence with 
the output of our actor-critic algorithm we are able to price and hedge arbitrary portfolios of derivatives.

Thirdly, we show how we can train multiple hedging strategies with different risk aversion levels in parallel, representing them all with a single 
neural network that learns the mapping from market state and risk aversion level to hedging action. We present this as a general actor-critic algorithm for solving 
risk-averse reinforcement learning problems across multiple risk aversion levels in Algorithm \ref{algo:actor_critic}.

\section{Reinforcement learning}
We first introduce the risk-averse reinforcement learning framework, before turning to our specific use-case of hedging portfolios of derivatives.

\subsection{Markov decision processes}
We consider the problem of sequential decision making of an agent acting within an environment over discrete timesteps over a finite time horizon. We formulate this as a Markov Decision Process $( \cal S,  \cal A, r, p)$, where at each timestep, 
the agent observes the current state $s_t \in \cal S$, takes an action $a_t \in \cal A$ based on the current state. Here we consider continuous state and action spaces: $\cal S \subset \R^d$, and similarly $\cal A \subset \R^n$. The transition dynamics are given by $p(s_{t+1} | s_t, a_t)$. The environment 
is in general stochastic, giving rise to the stochastic scalar-valued reward function $r = r(s_t, a_t), s_t \in \cal S, a_t \in \cal A$. The actions are drawn from a policy $\pi : \cal S \rightarrow \cal A$ which we will assume to be deterministic.        

It is standard to define the \emph{returns} from a state $s_t$ as the sum of future rewards from following the policy $\pi$ 

\eq{gains}
  R^\pi(s_t) =  \sum_{i=t}^{T} \,r(s_i,a_i) 
\eqend

Similarly, it is standard to define a value function over states as the expected returns of following the policy $\pi$ starting from state $s_t$

\eq{value}
  V^\pi(s_t) = \E \left[ R^\pi (s_t) | s_t \right]
\eqend

The objective of standard reinforcement learning is to learn a policy $\pi$ which maximizes the expected returns, typically from some distribution over initial states. In this formulation, only the expectation is taken into consideration, not the variance or higher moments of the distribution of future rewards. For this reason, we may describe this objective as \emph{risk neutral}. A
range of on-policy and off-policy algorithms exist, but all
approaches to solving this optimization make use of some
variation on the linear Bellman equation

\eq{linear_bellman}
V^\pi(s_t) = \E \left[ r(s_t, a^\pi_t ) + V^\pi(s_{t+1}) \right]
\eqend
and subsequently perform either value iteration, policy iteration or a combination of the two in actor-critic methods. See \cite{SUTTON, hambly2021recent, jaimungal2022reinforcement} for an overview of approaches.

\subsection{Risk-averse reinforcement learning}

When managing the risk of a portfolio of derivatives however, the inherent stochastic nature of the system makes
the variability of future
 cashflows of paramount importance, and an optimization criterion which accounts for both risk and reward is essential. 

To this end, we will not use expected returns, 
but a risk-adjusted return measure to assess the value of future rewards. We will assess our policy with respect to the certainty equivalent of the
\emph{exponential utility} which is given for any random variable ~$X$ as

\eq{entropy}
	U_\lambda( X ) = - \frac{1}{\lambda} \log \E \left[ \exp( - \lambda X) \right]
\eqend
for a risk-aversion parameter $\lambda \in \R^+$. The exponential utility has the limits ~$U(X) = \E[X]$ for $\lambda \rightarrow 0$ (risk-neutral)
and $U(X) = \inf X$ for~$\lambda \rightarrow \infty$ (worst-case). It is an example of a \emph{monetary utility} as it is increasing (more is better), 
concave (risk-averse) and cash-invariant, in the sense that for a fixed amount~$c \in \R$, we have $U(X+c) = U(X)+c$. 
This last property implies that ~$U_\lambda(X - U_\lambda(X)) = 0$, which means we may interpret $-U_\lambda(X)$
as the \emph{monetary utility equivalent} of~$X$ -- that is, the amount of cash we would need to add to the portfolio to make it acceptable for that level of risk aversion. 
The negative of the exponential utility is an example of a \emph{convex risk measure} \cite{follmer2010convex}, other examples include CVaR which has been explored in the risk-sensitive RL literature \cite{chow2014algorithms, chow2015risk, tamar2016sequential}, but has the somewhat unnatural property of 
\emph{coherence}: $U(nX) = nU(X)$ suggesting that the risk value of a position grows linearly with position size, whereas superlinear growth is more appropriate.   

The exponential utility has the Taylor expansion $U_\lambda(X) = \E[X] - \lambda/2 \mathrm{Var}[X] + \cal O(\lambda^2)$ and indeed when $X$ is normally distributed, it becomes equivalent to the classic Markowitz mean-variance portfolio objective. 

Applying the exponential utility to the MDP setting, we may define the value of a state now as the utility of the future returns from that state 

\eq{entropy_rl}
V^\pi(s_t) = U_\lambda \left( \sum_{i=t}^{T} \,r(s_i,a_i) \right)
\eqend
and the goal is to find a policy $\pi$ which maximizes this objective. In the original deep hedging approach, this was done relative to a fixed initial state $s_0$, and solved via periodic policy search \cite{DH}. Here we present a more general approach that is able to deal with a variety of initial states. 
It can be shown that the value function defined by \ref{eq:entropy_rl} satisfies a non-linear \emph{risk-averse} Bellman equation \cite{dowson2020multistage}

\begin{align}
\label{bellman_entropy}
V^\pi(s_t) &= U_\lambda \left ( r(s_t, a_t) + V^\pi(s_{t+1}) \right) \\
&= -\frac{1}{\lambda} \log \E \left[ \exp( - \lambda \left( r(s_t, a_t) + V^\pi(s_{t+1}) \right) \right] 
\end{align}

The exponential utility is the only utility apart from the simple expectation to satisfy this property of \emph{time consistency} \cite{gimelfarb2021risk, kupper2009representation}.

\subsection{Multiple levels of risk aversion}

The above formulation of the value function allows for sensitivity to risk, but leaves open the question of what a sensible level of risk aversion is. For low values of $\lambda$, our value 
function behaves almost like the expectation, and learned policies will be almost risk-neutral. On the other hand, for high $\lambda$, policies may be overcautious due to extreme sensitivity to large losses. Rather than choosing a specific level, we can consider an ensemble of agents, each with different risk aversion levels, where we want to learn an optimal policy for each agent. We can learn across these multiple agents in parallel, by including $\lambda$ in the ``agent state", hence the policy and the value function become implicitly dependent on the risk aversion level, i.e. $V^\pi(s_t ; \lambda)$ indicates the value of following policy $\pi$ from the state $s_t$, given that we apply the risk aversion $\lambda$ in the utility of returns. We drop any explicit dependence on $\lambda$ and still write $V^\pi(s_t)$ for brevity. 

We will consider risk aversion to be a static property of the agent which is independent of the environment state, and therefore does not change over time. Extensions to the case where the risk aversion is a function of the state $\lambda_t = \lambda(s_t)$ are possible, but we would lose the time consistency of the objective, and are not considered here.

Unlike the standard case of risk neutral reinforcement learning, it is not immediately clear how to use the risk-averse Bellman equation to perform policy or value updates. For the policy, note that the objective of $\sup_\pi V^\pi(s_t)$ is equivalent to solving the exponentiated objective

\eq{policy_objective}
\inf_\pi \frac{1}{\lambda} \exp \left( -\lambda V^\pi(s_t) \right) 
\eqend

Similarly, we can reformulate the Bellman equation for the value function as

\eq{dynamic_optimal}
\exp(-\lambda V^\pi(s_t)) = \E \left[ \exp \left( - \lambda \left( r(s_t, a_t) + V^\pi(s_{t+1}) \right) \right) \right] 
\eqend

Using these simple reformulations, we can develop an algorithm for solving the risk-averse MDP via actor-critic methods.

\subsection{Risk-averse actor-critic algorithm}
We now propose an actor-critic algorithm for solving this risk-averse MDP. We represent our policy and value functions with neural network 
function approximators, $\pi^\theta$ and $V^\phi$ respectively. By doing so, we can represent all policies across different levels 
of risk aversion by a single network, where the risk aversion level is included as a feature to the network, and do the same for the value function.

Then we can update the policy by following the gradient of the exponentiated objective with our 
approximated value function. That is, for the policy network we compute the gradient with respect to the loss

\eq{policy_loss}
\mathcal{L}(\theta) = \E \left[ \frac{1}{\lambda} \exp \left(- \lambda \left( r_t  +  V^\phi(s_{t+1}) \right) \right) \right]
\eqend
where the expectation is over states $s_t$ which are collected from our current policy.
For the critic, define the target as 

\eq{target}
y_t =  r_t + V^\phi(s_{t+1})
\eqend
then a natural candidate for the critic loss would be the squared error between the exponentiated value and the 
exponentiated target: $\E[ (e^{-\lambda y_t} - e^{-\lambda V^\phi(s_t)})^2]$. However, in practice this is observed to suffer from numerical instability -- 
any large, negative state value estimates can cause numerical issues, and the loss is notably asymmetric, i.e. for most 
states the loss can be decreased by simply increasing the estimates of the values of both the current and next state, irrespective of the observed reward, whilst also making the gradient close to zero. We are 
therefore relying on the corrective updates from end of episode value estimates where we have some ground truth such as $V(s_T) = 0$. This generally leads to very noisy and unstable updates. Instead, we construct a loss function which smooths the value function updates, and penalizes value function overestimates as well as underestimates, while still targeting the same 
exponentiated Bellman equation. So we consider the loss function 

\eq{critic_loss}
\mathcal{L}(\phi) = \E \left[ \frac{1}{\lambda} \exp \left(- \lambda  \left(y_t - V^\phi(s_t) \right) \right) - V^\phi(s_t)   \right]
\eqend
where again the expectation is over the states. The gradient of this loss with respect to the network weights is 

\eq{critic_grad}
\nabla_\phi \mathcal{L}(\phi) = \E \left[ \left( \exp \left(- \lambda \left(y_t - V^\phi(s_t) \right) \right) - 1 \right) \nabla_\phi V^\phi(s_t)  \right]
\eqend

The minimizer of this loss can be seen to satisfy the risk-averse Bellman equation, while the gradient no longer decays by simply increasing $V$. 
While the target $y_t$ depends on $\phi$, we follow the standard approach of ignoring this dependency and only updating the value function
with the gradients from $V^\phi(s_t)$. To do this, we use a copy of the network which we denote as the target network $\overline{V}$ which we do not train and copy the
parameters of $V^\phi$ over to this target. To stabilize the estimates of the target values, we use a soft update rule where the parameters described in \cite{lillicrap2015continuous}, which slowly transfers the weights via $\overline{\phi} = \tau \phi + (1 - \tau) \overline{\phi}$ for some small $\tau$.

By iteratively updating the policy and value functions on minibatches of experience sampled from the policy transitioning through the environment we obtain an actor-critic algorithm for the risk-averse MDP. We learn across risk aversion levels by specifying a set of levels of interest -- for example an interval $[\lambda_{min}, \lambda_{max}]$ -- and then sample from those at the beginning of each episode, assigning a different level to each trajectory in the minibatch. Since risk aversion tends to work on a logarithmic scale, we found that a good choice was sampling uniformly in log space. We summarize the full algorithm applied to episodes in Algorithm \ref{algo:actor_critic}. 

Note that (after some constant scaling and shifting), the loss for our critic is $\E [(y_t - V^\phi(s_t))^2] + \cal O(\lambda)$ so that 
in the limit $\lambda \rightarrow 0$ we recover the standard squared (linear) Bellman error used in risk-neutral algorithms. Similarly, the loss for our actor becomes $\E [ -r(s_t, a_t) - V^\phi(s_{t+1})]$, so that in the limit $\lambda \rightarrow 0$, our algorithm will learn the risk neutral policy.

\begin{algorithm}[htp]
    \caption{Multiple risk aversion actor-critic}\label{algo:actor_critic}
    \begin{algorithmic}
    \STATE {Initialize actor network $\pi$ and critic network $V$ with weights $\theta, \phi$}
    \STATE{Initialize target critic network $\overline{V}$ with weights $\overline{\phi} = \phi$}
    \FOR {Episode $1, \ldots, M$}
	\STATE {Sample a batch of $N$ initial states $s_0$ and risk aversion levels $\lambda$}
	\FOR {$t = 0, \ldots, T$} 
        		\STATE {Select actions from current policy $a_t = \pi(s_t)$}
        		\STATE {Execute actions,  receive rewards $r_t$ and transition into new states $s_{t+1}$}
		\STATE {Compute the target 

			$$  y_t = r_t + \overline{V}(s_{t+1}) $$   }

        		\STATE {Update the critic by gradient descent with respect to the loss 

	$$ \mathcal{L}(\phi) = \frac{1}{N} \sum_{i=1}^N \frac{1}{\lambda} \exp \left(- \lambda \left( y_t -  V(s_t)) \right) \right) - V(s_t) $$  }

    		\STATE {Update the actor by gradient descent with respect the loss
	
	$$ \mathcal{L}(\theta) = \frac{1}{N} \sum_{i=1}^N \frac{1}{\lambda} \exp \left(- \lambda \left( r_t + V(s_{t+1}) \right) \right) $$ }
		
		\STATE {Update target critic network $\overline{\phi} = \tau \phi + (1 - \tau) \overline{\phi} $ }	
	\ENDFOR
    \ENDFOR
    \end{algorithmic}
\end{algorithm}

\section{Deep Hedging}
We now focus on the primary application of our risk-averse actor-critic algorithm to solve the problem of hedging arbitrary portfolios of derivatives with respect to the 
exponential utility. 

\subsection{Market environment}
We consider a setting of trading in a financial market over discrete time steps.
We denote by $m_t$ the market state today. The market state contains all information available to
us today such as current market prices, time, past prices, bid/asks, social media feeds and so forth.

We assume that the agent is a trader who is in charge of a \emph{portfolio} -- also known as a ``book" -- of financial instruments such as securities, OTC derivatives or currencies, which we will denote by $z_t$, and the space of such portfolios $\cal Z$. This portfolio changes over time due to the changes in the market state, and as a result of the trader's activity. In order to risk manage her portfolio, the trader has access to $n$ liquid hedging instruments $h_t$ such as forwards, vanillas options, swaps. Across different market states they will usually not be the contractually same
fixed-strike, fixed-maturity instruments: instead, they will usually be defined relative the prevailing market in terms of time-to-maturities and strikes relative to at-the-money. See \cite{DH}
for details.

At each timestep the trader may trade in $a_t \in \R^n$ units of the liquid hedging instruments, which then become part of the portfolio at the next time step as follows. Denote by $z'_{t+1}$ the current portfolio $z_t$ at the next time step (i.e. as if no action were taken), and $h'_{t+1}$ the state at the next time step of the instruments available to trade at the current time step. Then the portfolio at the next time step will be $z_{t+1} = z'_{t+1} + a_t \cdot h'_{t+1}$. 

Trading incurs costs, which in general can depend on both the current portfolio and
the market. This allows modelling trading restrictions based on our current position such as short-sell
constraints, or restrictions based on risk exposure. We represent this by a transaction cost function $c(s, a)$
which is normalized to $c(s, 0) = 0$, non-negative, and convex, giving rise to a convex set of admissible actions $\cal A(s) = \{a \in \R^n: c(s, a) < \infty \}$. A simple example of such a cost function is proportional transaction cost $c(s_t, a_t) = \alpha|a_t||h_t|$. 

We assume that $p(m_{t+1} | s_t, a_t) = p(m_{t+1} | m_t)$ -- that is, actions have no market impact, and the future state of the market only depends on the current market state. 

At every time step, the portfolio produces random cashflows depending on the market state, which we will denote $r(z_t)$, and the trader must pay for any instruments traded, giving rise to the reward 

\eq{reward}
r(s_t, a_t) = r(z_t) - a_t \cdot h_t - c(s_t, a_t) 
\eqend 

It is common practice for human traders to consider the \emph{book value} $B(z)$ of their portfolio, which we define to be its official mark-to-market, computed using the prevailing
market data $m$. For liquidly traded instruments such as vanillas, this could be a simple closing price or a weighted mid-price. For exotics, this may be the result of
computing more complex derivative model prices. We could choose to include the change in book value of the portfolio $B(z_{t+1}) - B(z_t)$ inside the reward function, which has the benefit of reducing both the delay and sparsity of rewards -- for example, a 1M vanilla option has at most one cashflow, which is multiple timesteps after the action of trading it. Therefore by including the change in book value we could provide more reward signals to our agent. In this implementation choose to only use realized cashflows in the reward. However, we use the book value to address  any positions which are still outstanding at the end of the episode. At time $T$ we value the portfolio by its book value -- or equivalently, we enforce liquidation of the remaining portfolio at book value, so that $r_T = B(z_T)$. For example, if we have accumulated a position of $\delta$ units of an underlying $S$, then the book value at time $T$ could be $\delta S_T - c(s_T, \delta)$ -- the proceeds from liquidating the position at the current midprice.

\subsection{Representing portfolios}
The state can therefore be written as $s_t = (m_t, z_t)$. But as mentioned above, we need a representation of $z_t$ such that hedging instruments we have just traded can be added into our existing portfolio in a meaningful way. For this, we define a feature map $\Phi_t : \cal Z \rightarrow \R^d$ that maps any portfolio into a feature vector, $\Phi_t(z) = (\Phi^1_t(z), \ldots, \Phi^d_t(z))$. Moreover, let $\Phi_t$ be a \emph{Linear Markov} feature map, in the sense that the mapping of the combined portfolio $z_{t+1} = z'_{t+1} + a_t \cdot h'_{t+1}$ is given by

\eq{phi}
\Phi_{t+1}(z_{t+1}) = \Phi_{t+1} ( z'_{t+1}) + a_t \cdot \Phi_{t+1} (h'_{t+1}) 
\eqend

We then call $\Phi_t(z)$ a \emph{Linear Markov representation (LMR)} of the portfolio. 
To construct such a feature map, we follow the common practice of human traders of computing risk sensitivities -- or \emph{Greeks} -- of our portfolio instruments. The Greeks represent 
various sensitivities of the value of a derivative to a change in model parameters on which the financial instrument is dependent. For example, the delta of an instrument $z$ written on an underlying $S$ is defined as $\Delta = \partial z / \partial S$. Thus, we specify some model, and compute vectors of sensitivities for our portfolio and hedging instruments and use them to define the feature map $\Phi_t$. Note that the map changes at each step, since the sensitivities are computed with respect to the current model parameters, which are themselves conditional on the market state $m_t$. Importantly, we do not assume that the model used to compute the sensitivities perfectly reflects the true dynamics of the environment, only that there is sufficient information in the representation to learn policies from. The LMR can also include the book value and any other risk metrics which are linear in nature.

In addition, if $\cal Z$ represents some subset of portfolios with common, product-specific features, we can optionally include these in our feature representation, such as strikes, fixing dates and so on. These will not be part of the LMR but additional features passed to the networks.  

\subsection{Actor-critic deep hedging}
Having set up the reward function and state representation, we can apply Algorithm \ref{algo:actor_critic} to learn hedging strategies for arbitrary portfolios across a range of risk aversion levels. Using a market simulator, such as described in \cite{wiese2021multi}, we can randomly sample a batch of initial market states and portfolio states $s_0 = (m_0, z_0)$. Then we can transition through the simulated market, trading in the available hedging instruments and receiving rewards according to \ref{eq:reward}, and following the algorithm, update our actor and critic at each transition via stochastic gradient descent. Once trained, we will be able to hedge arbitrary portfolios that were seen in training, and from the monetary utility property, the value function enables us to price any portfolio, given the current market state.

\section{Numerical Results}

In this section we present some results in a simulated stochastic volatility environment to highlight the generality of our approach. For this purpose, we assume that our 
market follows the dynamics of the Heston \cite{heston1993closed} model, although we stress that this is only for illustration and our algorithm is model independent. we begin with a brief reminder of the Heston model. 

\subsection{Heston market environment}

Consider a market with a single underlying asset, with options written on that asset. The market dynamics under the Heston model are specified by the stochastic differential equations

\begin{align*}
\mathrm{d}S_t &= \mu S_t \mathrm{d}t  +  \sqrt{v}_t S_t \mathrm{d}B_t \\
\mathrm{d}v_t &= \kappa( \theta - v_t) \mathrm{d}t + \xi \sqrt{v}_t \mathrm{d}W_t
\end{align*}
where $B$ and $W$ are Brownian motions with correlation $\rho \in [-1, 1]$, $S$ is the liquidly tradeable asset with drift $\mu$, and $v$ is the stochastic variance 
process of $S$, which follows a mean reverting Cox-Ingersoll-Ross process, with mean reversion rate $\kappa$, mean reversion level $\theta$, and volatility of volatility $\xi$. 
Sample paths of $S$ are simulated using the Broadie-Kaya scheme \cite{andersen2010simulation, broadie2006exact}. We generate $200,000$ sample paths with parameters $\rho = -0.7, \mu = 0, \kappa = 8, \theta = 0.00625, \xi = 1$, with daily data over a finite time horizon of $T=30$ days and normalized to $S_0 = \$1$. The choice of $\mu = 0$ ensures that we have no \emph{statistical arbitrage} and so the policy is purely hedging \cite{drift}. For each path we randomly assign a short position in a call option with strike $K \in \{ 0.9, 0.925, \ldots, 1.1 \}$ expiring at $T$, with fixed notional of $\$100$, so that the payoff is $-100(S_T - K)^+$. For sample efficiency we keep the maturity of the portfolios constant, to avoid episodes where many trading steps are taken on an empty portfolio after the option has expired. We use proportional transaction costs of $0.002$.

For the LMR, we compute Black-Scholes Greeks for the portfolios, using the square root of the variance process as the implied volatility. Strictly speaking, this is unobserved, but the rationale is that we imagine a case where we are able to observe it through some other liquidly traded variance products. Furthermore, we do not compute Greeks with respect to the true, unknown, Heston model, but with respect to the Black-Scholes model, hence the feature vector is an imperfect representation of the state, in particular because the Black-Scholes model does not account for the stochasticity of volatility. The Greeks used for the LMR are delta ($\Delta$), gamma ($\Gamma$), vega ($\cal V$), theta ($\Theta$), vanna, charm, and vomma. These can all be computed via closed-form formulas in the Black-Scholes model. In addition to the LMR we add portfolio-specific features of strike and time-to-maturity. 

The use of the Greeks to represent our portfolio state presents the challenge of feature scaling -- for example the gamma of a close-to-expiry option can be orders of magnitude higher than its delta. Unlike in supervised learning, features cannot simply be rescaled in the naive way since we do not know a priori what the distribution of states under the optimal policy will be. Solutions to this problem have been explored, such as batch normalization \cite{ioffe2015batch}, which employs dynamic feature scaling, using running means and standard deviations of the states. Instead of this, we take a policy which we know to be reasonably close to the optimal policy and compute the state mean and standard deviations under this policy for feature normalization. In this case, we can do so by running the Black-Scholes delta hedging policy on the environment $\pi^{BS}(s_t) = -\Delta(s_t)$. We will also use this policy as a benchmark for performance comparison.

For the actor, we use a three layer feedforward neural network with 256 units in each layer, and sigmoid activations. We use a skip connection, so that the output of the network is $\pi(s_t) = -\Delta(s_t) + \pi^\theta(s_t)$, which can be interpreted as learning the correct adjustment to the Black-Scholes delta hedge.

For the critic, we also use a three layer feedforward neural network with 256 units in each layer, and sigmoid activations. A reasonable first order approximation of the state value should be the book value of our current portfolio. Hence, our value function outputs $V(s_t) = B(s_t) + V^\phi(s_t)$. This can be seen as making the correct risk adjustment to the model implied cash value of our current portfolio. For the soft target updates we used $\tau = 0.001$.

At the beginning of each episode, we sample a batch of risk aversions uniformly on a log scale from the interval $[10^{-4}, 1]$, and the risk aversion for each trajectory is held fixed through the episode. Thus the actor network will learn hedging policies across these risk aversion levels. 

The learning rates were $1 \times 10^{-3}$ for the actor and $1 \times 10^{-4}$ for the critic. We train with minibatch sizes of $2048$ for $100,000$ episodes.

\subsection{Assessing the value function}
To assess the performance of the critic, we check that it correctly matches the true realized utility by generating $100$ random portfolios, with strike sampled uniformly from $[0.9, 1.1]$ and $\lambda$ sampled uniformly from $[10^{-4}, 1]$. We then compute the true value of each portfolio by running the policy on a sample of $10,000$ paths and then computing the utility of the terminal hedged PnL. We then compute the RMSE between value function estimates and true utility, giving a value of $RMSE= 0.014$ indicating that our value function is able to almost perfectly reproduce the correct utility.

\begin{figure}[htp]
   \centering 
    \includegraphics[width=0.5\textwidth]{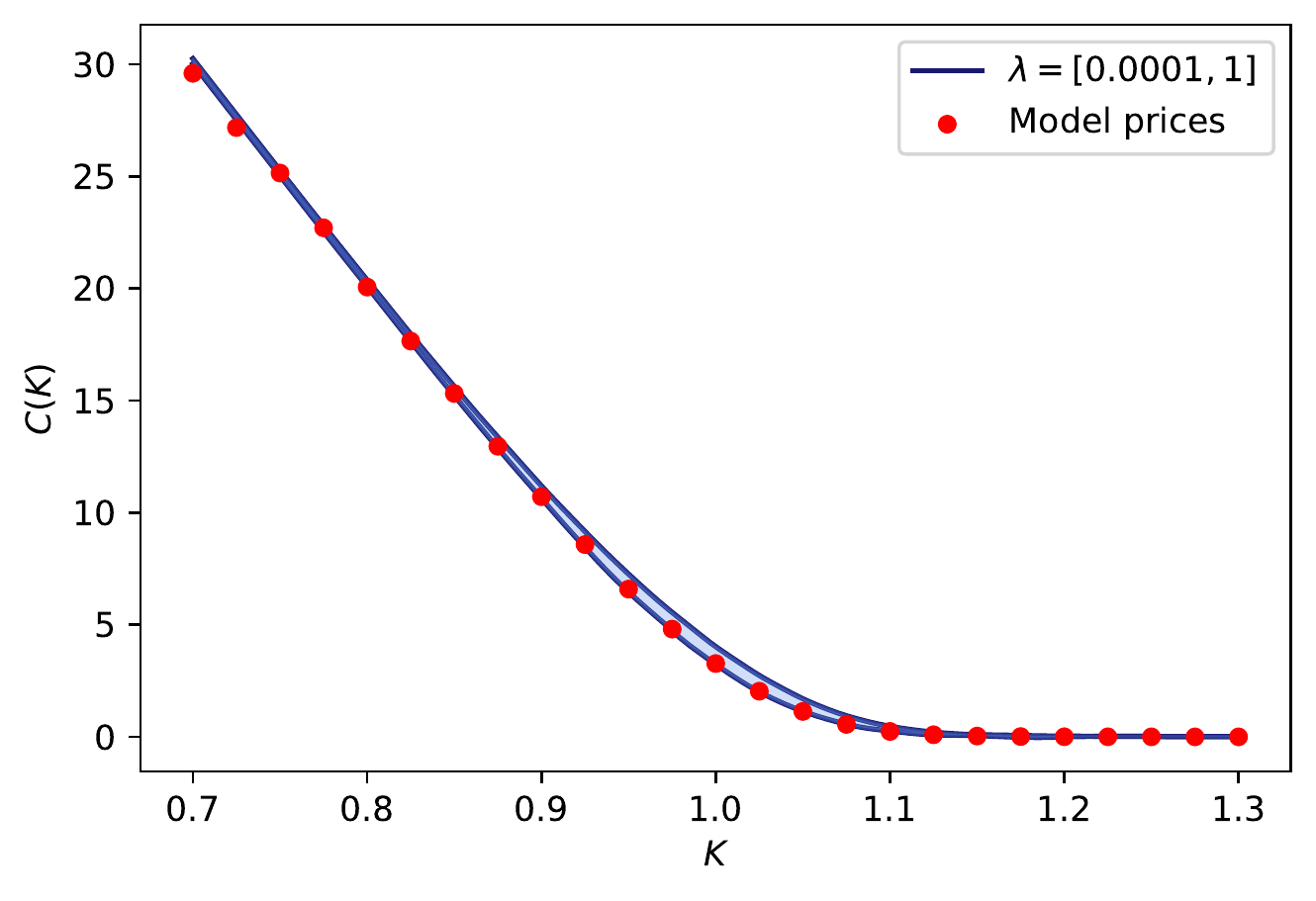}
    \caption{Call prices obtained by the value function for maturity $T=30$ compared with risk-neutral model prices. Low $\lambda$ produces 
prices very close to the model price, and higher $\lambda$ results in additional risk premium added.}
    \label{fig:call_prices_more_strikes}
\end{figure}

Next, we check that the value assigned to the portfolio corresponds correctly to the risk-adjusted price, as implied by the monetary utility. In Figure \ref{fig:call_prices_more_strikes} we compare the prices of $30$-day call options over a range of risk aversions. As $\lambda \rightarrow 0$ the value function correctly approximates the risk neutral model price for low $\lambda$, and since  $U_\lambda(X) \leq \E[X]$, as $\lambda$ increases we charge an additional risk premium to make the trade acceptable. Note that to get this correspondence it was essential that the market was free from statistical arbitrage, in which the value of the empty portfolio is zero. We also achieve very good extrapolation performance, pricing strikes outside of the range $[0.9, 1.1]$ seen in the training data with high degree of accuracy.

\begin{figure}[htp]
   \centering
    \includegraphics[width=0.5\textwidth]{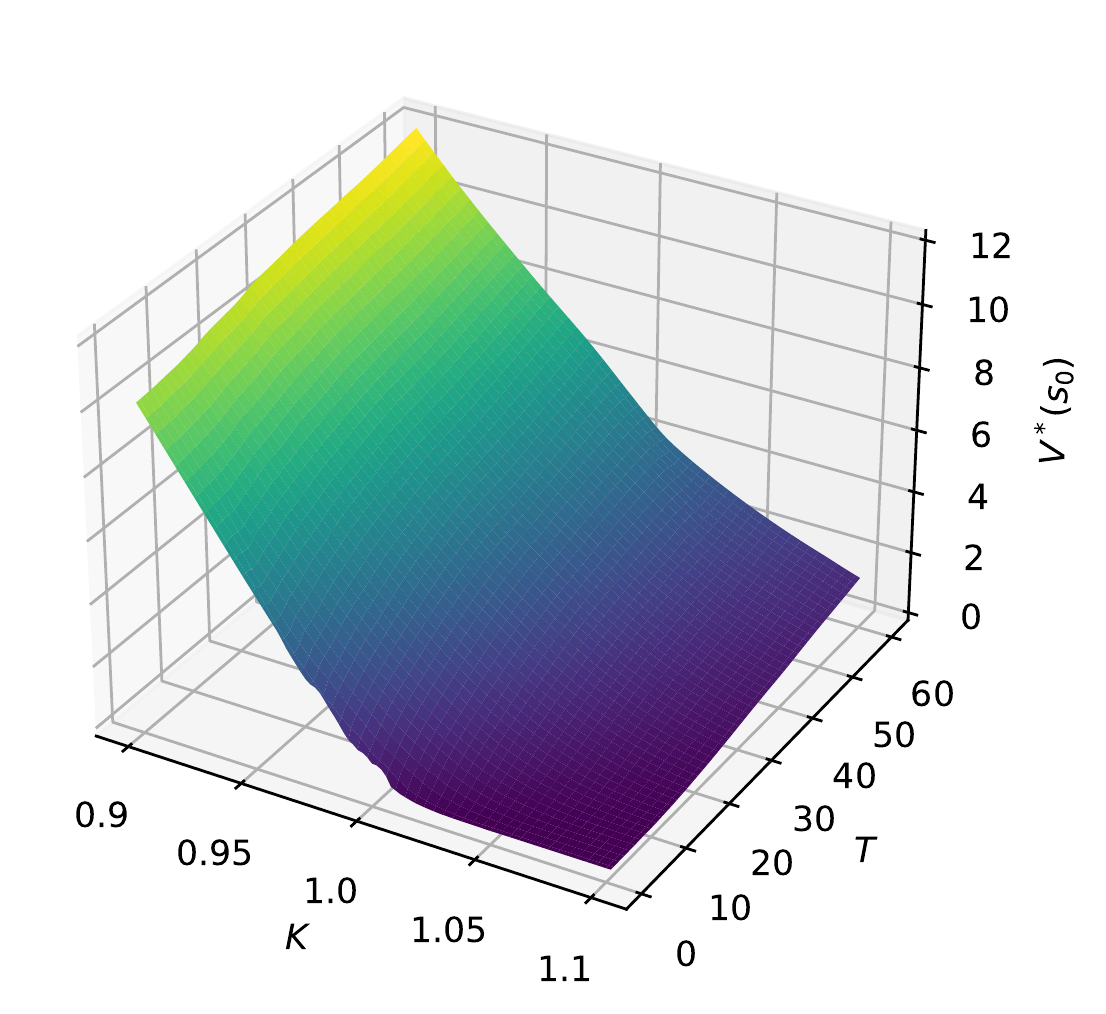}
    \caption{Value function estimates of call prices for maturities between $1-60$ days. Maturities from $31-60$ were never seen in training.}
    \label{fig:call_surface}
\end{figure}

Finally, in Figure \ref{fig:call_surface} we price options for maturities in the range $1-60$ days. We observe that the value function extrapolates well, preserving the convexity in strike and monotonicity in maturity, despite no explicit constraints. 

\subsection{Assessing the policy}

\begin{figure}[htp]
    \centering 
    \includegraphics[width=0.5\textwidth]{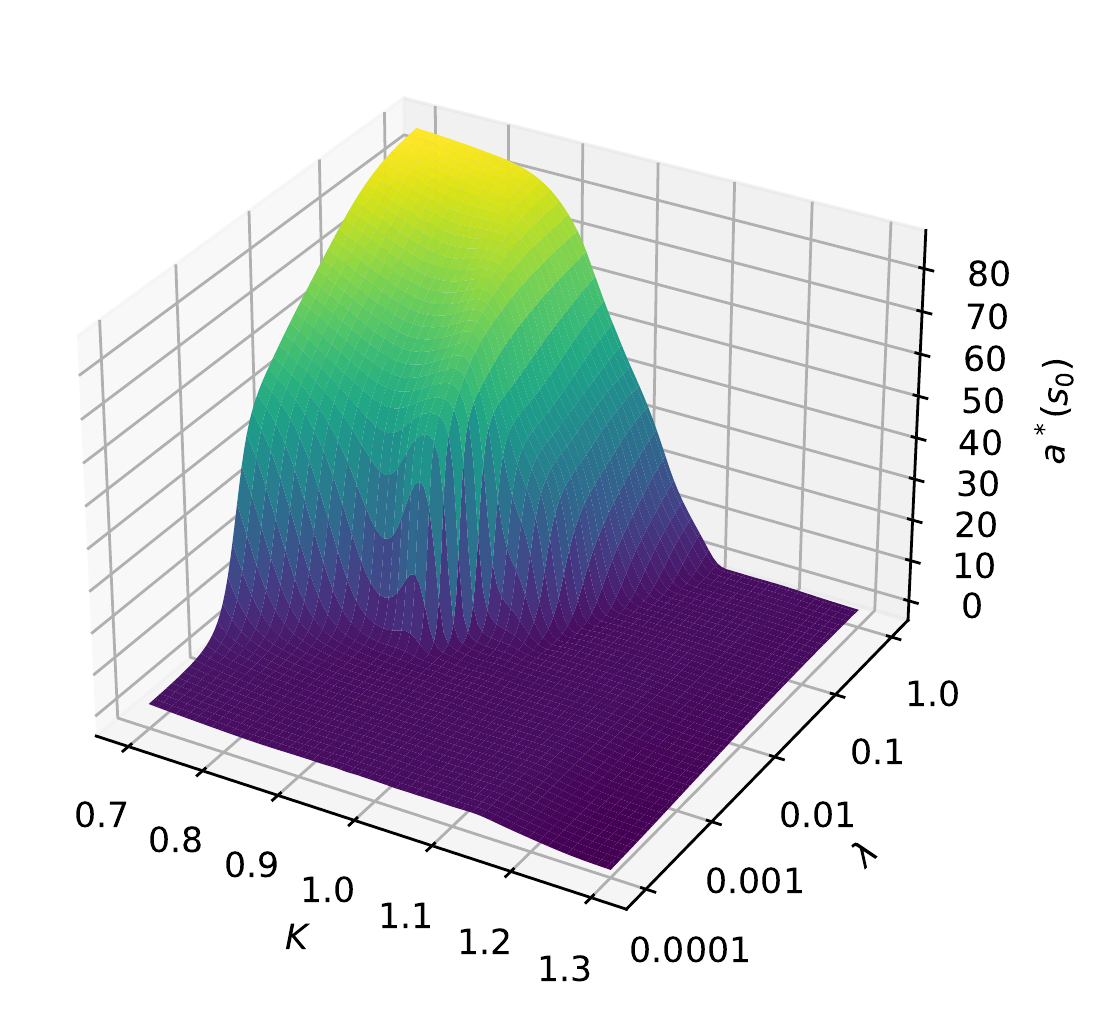}
    \caption{Initial hedge as a function of strike and risk aversion. The policy network smoothly interpolates from zero position for low $\lambda$ to an almost full hedge for high $\lambda$.}
    \label{fig:actions}
\end{figure}

We now assess the performance of the hedging policy. Figure \ref{fig:actions} shows the initial hedging action as a function of strike and risk aversion. For low $\lambda$, the policy leaves all positions unhedged. As we increase the risk aversion, the level of hedging smoothly increases. For deep out of the money options, the agent prefers to leave it unhedged even at higher risk aversion levels. Due to market frictions, the value of the unhedged position is higher than the hedged one.

\begin{figure}[htp]
    \centering 
    \includegraphics[width=0.5\textwidth]{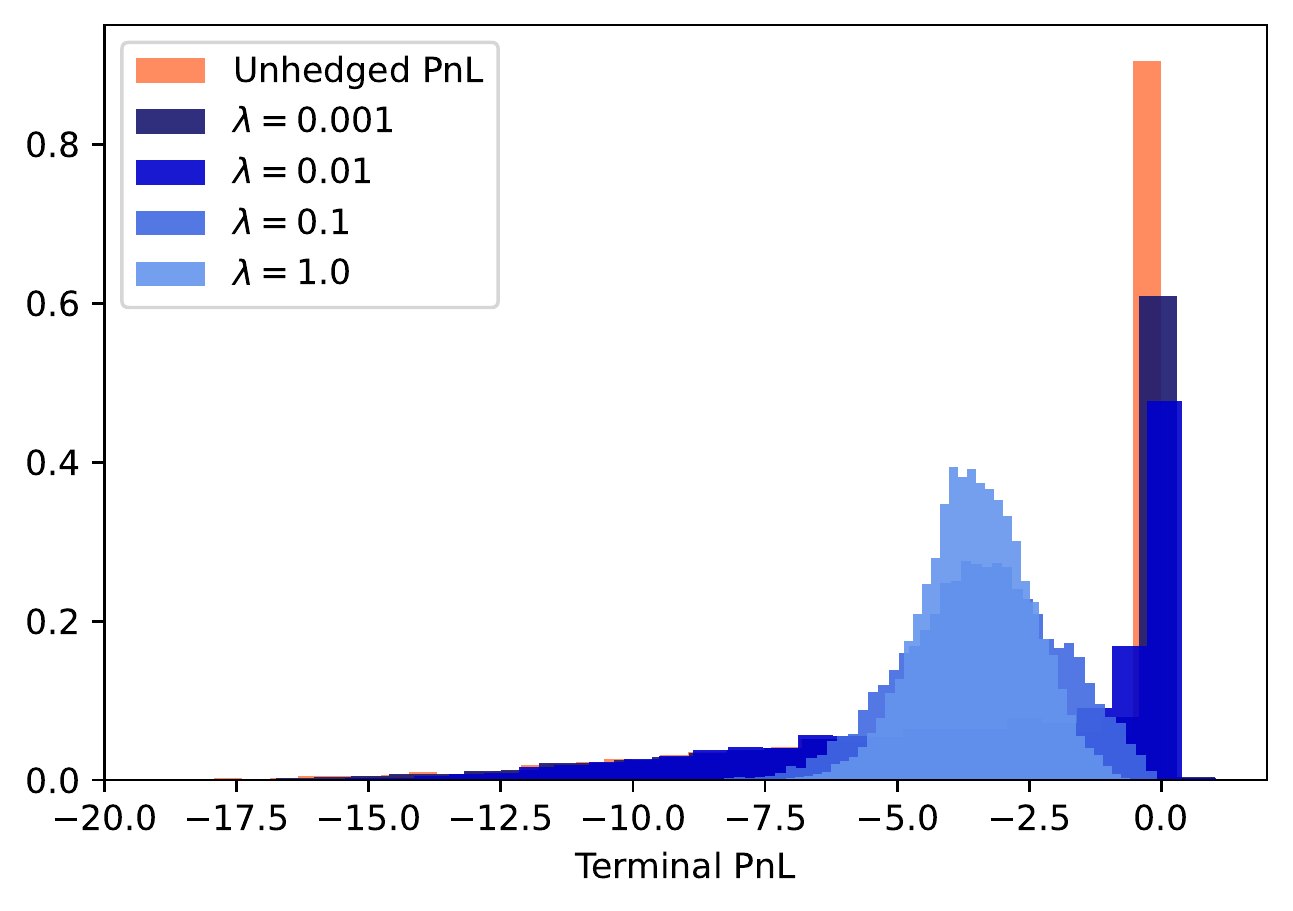}
    \caption{Terminal hedged PnL of a 30-day at the money call option for varying risk aversion levels.}
    \label{fig:terminal_pnl}
\end{figure}

Figure \ref{fig:terminal_pnl} compares the distribution of terminal hedged PnL of the policy at different risk aversion levels, showing tightening of the distribution as risk aversion increases. Note also that by the cash invariance of the value function, each distribution is centred around the (negative) corresponding risk adjusted price.

In Table \ref{tab:results} we compare our actor-critic method with the baselines of Black-Scholes delta hedging and vanilla deep hedging across a range of strikes for $\lambda = 0.1$. For vanilla deep hedging we train a new model for each payoff and then test the utility of all policies on a test set of $20,000$ new paths. The vanilla deep hedging model used the same network architecture as the policy network and trained for $1,000$ epochs. The policy reproduces the utility of vanilla deep hedging, which is in all cases better than delta hedging. 

 \begin{table}
\begin{tabular}{P{0.8cm} | P{2.2cm} P{2.2cm} P{1.8cm} } 

 Strike & Black-Scholes delta hedging & Vanilla deep hedging & Actor-critic deep hedging  \\ 
 \hline
$ 0.9 $ & $\minus 11.10$ & $\minus 10.87 $ & $\minus 10.91 $ \\ 
 
 0.95 & $ \minus 7.06$ & $\minus 6.74 $ & $\minus 6.68 $ \\

 1.0 & $\minus 3.78$ & $\minus 3.36$ & $\minus 3.41$ \\
 
 1.05 & $\minus 1.52$ & $\minus 1.14 $ & $\minus 1.11 $ \\
 
1.1 & $\minus 0.43$ & $\minus 0.26 $ & $\minus 0.27$    \\
\end{tabular}
\vspace{5mm}
  \caption{Utility of hedged PnL against strike, for the Black-Scholes delta hedging baseline, vanilla deep hedging and our actor-critic approach.}
\label{tab:results}
\end{table}

\begin{figure}[htp]
    \centering 
    \includegraphics[width=0.5\textwidth]{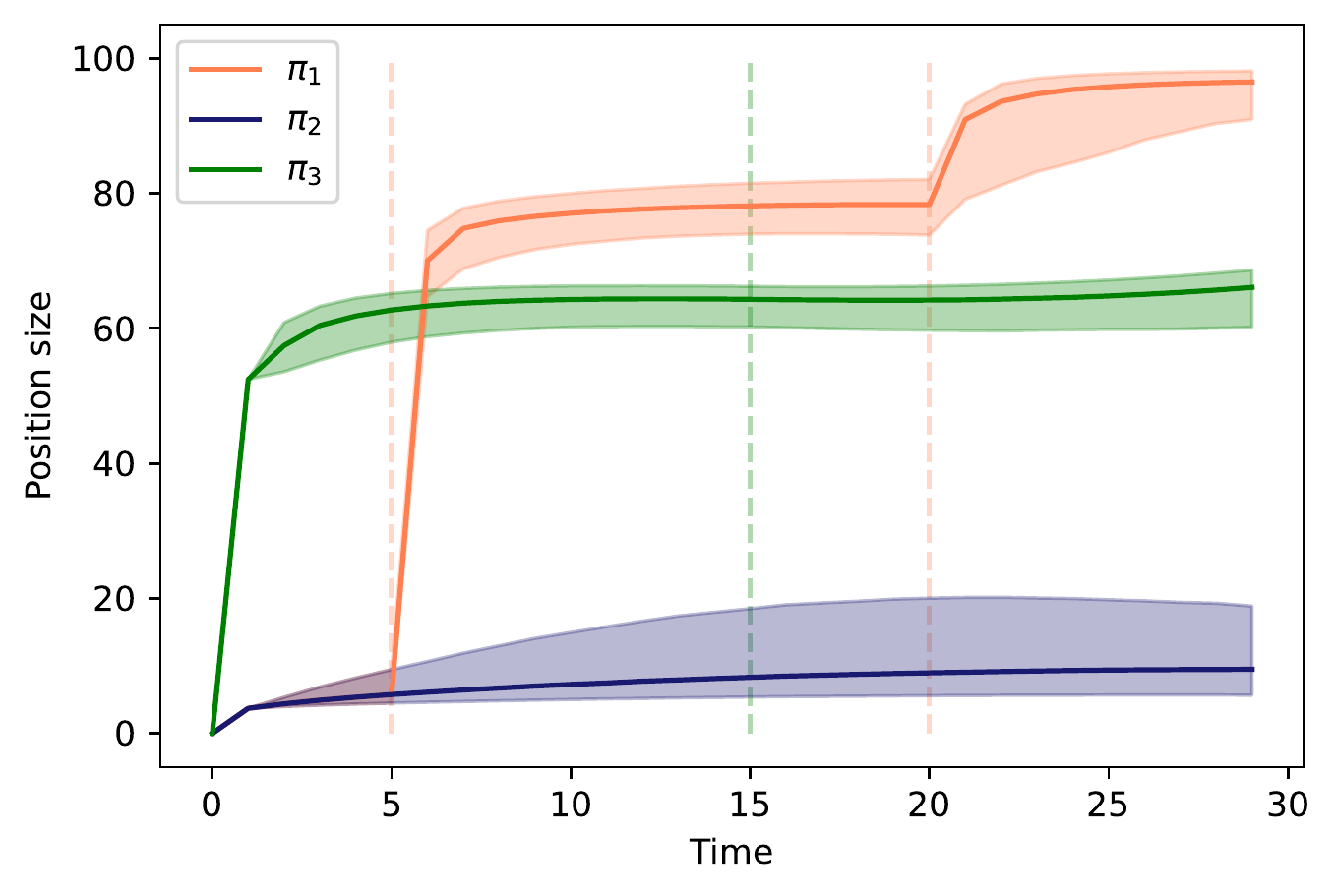}
    \caption{Plot of median and quartiles of hedge positions under dynamically adjusted risk aversion level. Under policy $\pi_1$ the agent increases its risk aversion at $t=5$ and again at $t=20$, both times leading to an instant increase in the hedge position.}
    \label{fig:hedge_policies}
\end{figure}

To illustrate the effect of dynamically changing the risk aversion level, we test the policy on an environment with a position in a $1.1$ strike call. $\pi_1$ has $\lambda = 0.005$ for all $t$, $\pi_2$ starts with $\lambda = 0.005$ but then increases to $\lambda = 0.1$ at $t=5$ and then $\lambda = 1$ at $t = 20$, and $\pi_3$ starts with $\lambda = 0.01$, then decreases to $\lambda = 10^{-4}$ at $t=15$. Figure \ref{fig:hedge_policies} shows the median and quartiles of the hedging positions as we evaluate on the environment. Each time the risk aversion increases, the policy instantly increases its hedge, but when the risk aversion drops, it prefers to just keep the position until maturity, to avoid paying transaction costs.

\section{Conclusion}
In this paper, we proposed an actor-critic algorithm for solving risk-averse stochastic control problems in continuous state and action spaces, across multiple risk aversion levels, using neural networks to approximate both the policy and a value function representing an optimized certainty of the exponential utility. With this algorithm, we were able to learn prices and hedging strategies for a range portfolios of derivatives across multiple risk aversion levels in a market with stochastic volatility. We utilised a Linear Markov representation of the portfolio state, and our policy was able to learn the correct state-dependent adaptations to model parameters to give an optimal hedging policy.
This paper therefore lays the groundwork for a full ``Bellman" deep hedging, and future work extending to the infinite time horizon setting and expanding the numerical implementation to a broader class of exotics and hedging instruments will be explored.

\bibliographystyle{ACM-Reference-Format}
\bibliography{references}

\section*{Disclaimer}
Opinions and estimates constitute our judgement as of the date of this Material, are for informational purposes only and are subject to change without notice. It is not a research report and is
not intended as such. Past performance is not indicative of future results. This Material is not the
product of J.P. Morgan’s Research Department and therefore, has not been prepared in accordance
with legal requirements to promote the independence of research, including but not limited to, the
prohibition on the dealing ahead of the dissemination of investment research. This Material is not
intended as research, a recommendation, advice, offer or solicitation for the purchase or sale of
any financial product or service, or to be used in any way for evaluating the merits of participating
in any transaction. Please consult your own advisors regarding legal, tax, accounting or any other
aspects including suitability implications for your particular circumstances. J.P. Morgan disclaims
any responsibility or liability whatsoever for the quality, accuracy or completeness of the information herein, and for any reliance on, or use of this material in any way. Important disclosures at:
www.jpmorgan.com/disclosures.

\end{document}